\documentclass[journal]{IEEEtran}
\usepackage{graphicx}
\usepackage{color}
\usepackage{placeins}
\usepackage{float}
\usepackage{tabularx,colortbl}
\usepackage{amssymb}
\usepackage{amsthm}
\usepackage{cite}
\usepackage{amsmath}
\usepackage{makecell}
\usepackage{subfigure}

\theoremstyle{plain}
\newtheorem{thm}{Theorem}

\newtheorem{coro}{Corollary}

\theoremstyle{plain}
\newtheorem{rem}{Remark}

\IEEEoverridecommandlockouts

\begin{document}


\title{Optimal Bilinear Equalizer Beamforming Design for Cell-Free Massive MIMO Networks with Arbitrary Channel Estimators
\thanks{Z. Wang, J. Zhang, H. Lei, and B. Ai are with the School of Electronics and Information Engineering, Beijing Jiaotong University, Beijing 100044, P. R. China (e-mail: \{zhewang\_77, jiayizhang, haolei, boai\}@bjtu.edu.cn). D. Niyato is with the College of Computing \& Data Science, Nanyang Technological University, Singapore 639798 (e-mail: dniyato@ntu.edu.sg).}}
\author{Zhe Wang, Jiayi Zhang,~\IEEEmembership{Senior Member,~IEEE,} Hao Lei, Dusit Niyato,~\IEEEmembership{Fellow,~IEEE}, Bo Ai,~\IEEEmembership{Fellow,~IEEE}\vspace{-1.1cm}}
\maketitle

\begin{abstract}
This paper studies the distributed optimal bilinear equalizer (OBE) beamforming design for both the uplink and downlink cell-free massive multiple-input multiple-output networks. We consider arbitrary statistics-based channel estimators over spatially correlated Rician fading channels. In the uplink, we derive the achievable spectral efficiency (SE) performance and OBE combining schemes with arbitrary statistics-based channel estimators and compute their respective closed-form expressions. It is insightful to explore that the achievable SE performance is not dependent on the choice of channel estimator when OBE combining schemes are applied over Rayleigh channels. In the downlink, we derive the achievable SE performance expressions with BE precoding schemes and arbitrary statistics-based channel estimators utilized and compute them in closed form. Then, we obtain the OBE precoding scheme leveraging insights from uplink OBE combining schemes. 

\end{abstract}

\IEEEpeerreviewmaketitle
\vspace*{-0.4cm}
\section{Introduction}
Multiple-input multiple-output (MIMO) technology has been a promising empowered technology for wireless communication networks since forth-generation (4G) wireless communication networks \cite{5595728,hong2024robust,ZheSurvey,wu2024exploit}. Cell-free massive MIMO (CF mMIMO) technology has been extensively studied, which is viewed as one the most important technologies for future wireless communication networks \cite{7827017,[162],buzzi2019user}. In CF mMIMO networks, a large number of access points, which are connected to one or several central processing units (CPUs), provide the uniform service to user equipment (UEs) \cite{7827017}. Various processing schemes for CF mMIMO networks, differing in the level of CPU involvement and the degree of assistance among APs, can be implemented \cite{[162]}.

Beamforming design is very vital in CF mMIMO networks to fully realize their potential. The centralized processing scheme employing the minimum mean-square error (MMSE) combining is recognized as the most competitive processing scheme \cite{[162]}, but it embraces very significant computational complexity due to the instantaneous channel state information (CSI)-based matrix inversion. To alleviate the burden of computational complexity, while maintaining excellent performance, several works \cite{polegre2021pilot,OBETrans} have been studied to design optimal bilinear equalizer (OBE)-based beamforming schemes based on the fundamentals in \cite{neumann2018bilinear}. The bilinear equalizer (BE) denotes a series of schemes designed by the product of the statistics-based matrix and the instantaneous channel state information. The OBE scheme can be derived by optimizing the BE matrix to maximize the rate performance. The authors in \cite{polegre2021pilot} investigated OBE beamforming schemes, which were called ``generalized maximum ratio" schemes, over Rayleigh fading channels-based CF mMIMO systems with both APs and UEs equipped with single antenna. The authors in \cite{OBETrans} considered a more generalized CF mMIMO network over Rician fading channels with multi-antenna APs. One centralized and two distributed uplink (UL) OBE combining schemes were investigated in \cite{OBETrans}.

According to \cite{OBETrans}, the channel statistics pose significant impacts on the achievable performance for the OBE schemes. Among representative channel statistics in CF mMIMO systems, the channel estimator-related covariance matrices showcase vital importance. Thus, it is important to study the OBE beamforming schemes with more generalized channel estimators than the MMSE one and explore the effects of channel estimators on achievable performance. However, the OBE beamforming schemes in \cite{polegre2021pilot,OBETrans} are all implemented based on the MMSE channel estimator, which are not suitable for scenarios that utilize channel estimators other than the MMSE estimator. Meanwhile, only the UL OBE combining schemes were studied in \cite{OBETrans} and it is also insightful to investigate the downlink (DL) OBE precoding schemes. Motivated by the above motivations, in this paper, we investigate the OBE beamforming design for both the UL and DL in CF mMIMO systems with Rician fading channels with arbitrary statistics-based channel estimators and multi-antenna APs. The major contributions are summarized as follows.
\begin{itemize}
\item In the UL, we propose the OBE combining scheme with arbitrary statistics-based channel estimators over Rician channels. Closed-form expressions for achievable spectral efficiency (SE) and OBE combining expressions are derived. Crucially, it is insightful to find that, with OBE combining applied, arbitrary statistics-based channel estimators would achieve the same achievable SE performance over Rayleigh channels.
\item In the DL, achievable SE expressions with BE precoding utilized and their respective closed-form expressions with arbitrary statistics-based channel estimators applied are derived. Relying on the channel reciprocity, we derive the OBE precoding schemes, which also hold for arbitrary statistics-based channel estimators. 
\item In numerical results, we observe that OBE beamforming schemes outperform the regularized zero-forcing-based processing schemes. Meanwhile, we validate the correctness of derived analytical results and confirm the insightful conclusion that the UL achievable SE performance with OBE combining applied remains unaffected by the choice of the channel estimator.
\end{itemize}

\textbf{\emph{Notation}}: We denote column vectors and matrices as boldface lowercase letters $\mathbf{a}$ and boldface uppercase letters $\mathbf{A}$. Let $\left( \cdot \right) ^H$ and $\left( \cdot \right) ^T$ denote the conjugate transpose and transpose. $\mathbf{a}\sim \mathcal{N} _{\mathbb{C}}\left( 0,\mathbf{R} \right)$ is a complex Gaussian distribution vector with $\mathbf{R}$ being the correlation matrix. We define $\mathbb{E} \left\{ \cdot \right\}$ and $\mathrm{tr}( \cdot )$ as the expectation and trace operators, respectively. $\mathrm{vec}\left( \cdot  \right)$ denotes the vectorization operator and $\mathbf{I}_{N\times N}$ denotes the $N\times N$ identity matrix. We define the Kronecker product and Euclidean norm as $\otimes$ and $\left\| \cdot \right\|$, respectively.

\vspace*{-0.4cm}

\section{System Model}\label{fundamentals}
\vspace*{-0.1cm}

We investigate a CF mMIMO system, which is composed of $M$ APs with $N$ antennas for each AP, connected via fronthaul links to a CPU, and $K$ UEs with a single antenna each. Each time-frequency coherence block consists of $\tau_c$ channel uses. We apply the standard massive MIMO time-division duplex (TDD) mode \cite{8187178}, where $\tau_p$, $\tau_u$, and $\tau_d$ channel uses are reserved for the UL pilot transmission, UL data transmission, and DL data transmission, respectively, that is $\tau_c=\tau_p+\tau_u+\tau_d$. The channel response between AP $m$ and UE $k$ can be denoted as $\mathbf{g}_{mk}\in \mathbb{C} ^N$, which stay constant in each coherence block. 
The spatially correlated Rician fading channel is considered as $\mathbf{g}_{mk}=\bar{\mathbf{g}}_{mk}+\check{\mathbf{g}}_{mk}\in \mathbb{C} ^N$, where $\bar{\mathbf{g}}_{mk}$ and $\check{\mathbf{g}}_{mk}\sim \mathcal{N} _{\mathbb{C}}( \mathbf{0},\check{\mathbf{R}}_{mk} ) $ denote the deterministic LoS and stochastic NLoS components \cite{pan2023joint}, respectively, with $\check{\mathbf{R}}_{mk}\in \mathbb{C} ^{N\times N}$ being the correlation matrix.

Then, we focus on the channel estimation. We apply $\tau_p$ mutually orthogonal pilot signals, $\boldsymbol{\phi} _1,\dots ,\boldsymbol{\phi} _{\tau _p}$, with $\tau_p$-length each and $\left\| \boldsymbol{\phi} _t \right\| ^2=\tau _p$. We consider a practical scenario with the pilot contamination, where a same pilot signal can be allocated to more than one UE. We let $\mathcal{P} _k$ be the subset of UEs that apply the same pilot signal as UE $k$ $\boldsymbol{\phi} _{t_{k}}$ including itself. All UEs simultaneously send the pilot and the received pilot signal after despreading at AP $m$ for UE $k$ can be formed as \cite{8187178} $\mathbf{y}_{mk}^{p}=\sum_{l\in \mathcal{P} _k}{\sqrt{p_l}\tau _p\mathbf{g}_{ml}}+\mathbf{n}_{mk}^{p}$, where $p_l$ represents the UL transmitted power of UE $l$ and $\mathbf{n}_{mk}^{p}\sim \mathcal{N} _{\mathbb{C}}( \mathbf{0},\tau _p\sigma ^2\mathbf{I}_N ) $ is the thermal noise with $\sigma ^2$ being the noise power. Relied on $\mathbf{y}_{mk}^{p}$, the estimate of $\mathbf{g}_{mk}$ can be derived based on different criteria. One well-studied channel estimator is the MMSE estimator \cite{enyusurvey} as $\hat{\mathbf{g}}_{mk}^{\mathrm{MMSE}}=\bar{\mathbf{g}}_{mk}+\sqrt{p_k}\check{\mathbf{R}}_{mk}\mathbf{\Psi }_{mk}^{-1}\left( \mathbf{y}_{mk}^{p}-\bar{\mathbf{y}}_{mk}^{p} \right)$, where $\bar{\mathbf{y}}_{mk}^{p}=\sum_{l\in \mathcal{P} _k}{\sqrt{p_l}\tau _p\bar{\mathbf{g}}_{ml}}$ and $\mathbf{\Psi }_{mk}=\sum_{l\in \mathcal{P} _k}{p_l\tau _p\check{\mathbf{R}}_{ml}}+\sigma ^2\mathbf{I}_N$. Motivated by the structure of $\hat{\mathbf{g}}_{mk}^{\mathrm{MMSE}}$, we study a generalized channel estimator as
\begin{equation}\label{CE}
\hat{\mathbf{g}}_{mk}=\bar{\mathbf{g}}_{mk}+\mathbf{A}_{mk}( \mathbf{y}_{mk}^{p}-\bar{\mathbf{y}}_{mk}^{p}), 
\end{equation}
where $\mathbf{A}_{mk}\in \mathbb{C} ^{N\times N}$ is an arbitrary statistic information-based matrix, which does not influence the distribution of $\mathbf{g}_{mk}$. Thus, we have $\hat{\mathbf{g}}_{mk}\sim \mathcal{N} _{\mathbb{C}}( \bar{\mathbf{g}}_{mk},\tau _p\mathbf{A}_{mk}\mathbf{\Psi }_{mk}\mathbf{A}_{mk}^{H} ) $. For the channel estimation error $\tilde{\mathbf{g}}_{mk}=\mathbf{g}_{mk}-\hat{\mathbf{g}}_{mk}$, we have $\tilde{\mathbf{g}}_{mk}\sim \mathcal{N} _{\mathbb{C}}( \mathbf{0},\mathbf{C}_{mk}) $ with $\mathbf{C}_{mk}=\check{\mathbf{R}}_{mk}-\sqrt{p_k}\tau _p\check{\mathbf{R}}_{mk}\mathbf{A}_{mk}^{H}-\mathbf{A}_{mk}\sqrt{p_k}\tau _p\check{\mathbf{R}}_{mk}+\tau _p\mathbf{A}_{mk}\mathbf{\Psi }_{mk}\mathbf{A}_{mk}^{H}$.

\begin{rem}
$\mathbf{A}_{mk}$ in \eqref{CE} can be modelled based on various estimation criteria and practical applications. For instance, $\mathbf{A}_{mk}=\sqrt{p_k}\check{\mathbf{R}}_{mk}\mathbf{\Psi }_{mk}^{-1}$ denotes the MMSE estimator as \cite[Eq. (1)]{OBETrans}. $\mathbf{A}_{mk}=\sqrt{p_k}\hat{\mathbf{R}}_{mk}\hat{\mathbf{\Psi}}_{mk}^{-1}$ denotes the approximate MMSE estimator with imperfect channel covariance information as in \cite[Eq. (13)]{bjornson2016massive}, where $\hat{\mathbf{R}}_{mk}$ and $\hat{\mathbf{\Psi}}_{mk}$ denotes the imperfect representations of $\check{\mathbf{R}}_{mk}$ and $\mathbf{\Psi }_{mk}$, respectively. Meanwhile, when $\mathbf{A}_{mk}=1/( \sqrt{p_k}\tau _p ) \mathbf{I}_N$, we define this estimator as the generalized least-square (GLS) estimator. The GLS estimator extends the conventional LS estimator as indicated in \cite[Eq. (15)]{ozdogan2019massive}, by specifically extracting the deterministic LoS component during the estimation.
\end{rem}
\vspace*{-0.4cm}

\newcounter{mytempeqncnt1}
\begin{figure*}[t]
\normalsize
\setcounter{mytempeqncnt1}{\value{equation}}
\setcounter{equation}{1}
\begin{align} \label{SINR_ul}
\mathrm{SINR}_{k}^{\mathrm{ul}}=\frac{p_k| \sum_{m=1}^M{\mathbb{E} \{ \mathbf{v}_{mk}^{H}\mathbf{g}_{mk} \}} |^2}{\sum_{l=1}^K{p_l\mathbb{E} \{ | \sum_{m=1}^M{\mathbf{v}_{mk}^{H}\mathbf{g}_{ml}} |^2 \}}-p_k| \sum_{m=1}^M{\mathbb{E} \{ \mathbf{v}_{mk}^{H}\mathbf{g}_{mk} \}} |^2+\sigma ^2\sum_{m=1}^M{\mathbb{E} \{ \| \mathbf{v}_{mk} \| ^2 \}}}
\end{align}
\setcounter{equation}{\value{mytempeqncnt1}}
\hrulefill
\vspace*{-0.6cm}
\end{figure*}

\newcounter{mytempeqncnt2}
\begin{figure*}[t]
\normalsize
\setcounter{mytempeqncnt2}{\value{equation}}
\setcounter{equation}{2}
\begin{align} \label{SINR_Distributed_Closed}
\overline{\mathrm{SINR}}_{k}^{\mathrm{ul}}=\frac{p_k| \sum_{m=1}^M{\mathrm{tr}( \mathbf{W}_{mk}^{H}\bar{\mathbf{R}}_{mk} )} |^2}{\left( \begin{array}{c}
	\sum_{l=1}^K{\sum_{m=1}^M{p_l\varepsilon _{mkl}}}+\sum_{l\in \mathcal{P} _k}{\sum_{m=1}^M{p_l\xi _{mkl}}}+\sum_{l=1}^K{p_l( | \sum_{m=1}^M{\chi _{mkl}} |^2-\sum_{m=1}^M{| \chi_{mkl} |^2} )}\\
	-p_k| \sum_{m=1}^M{\mathrm{tr}( \mathbf{W}_{mk}^{H}\bar{\mathbf{R}}_{mk} )} |^2+\sigma ^2\sum_{m=1}^M{\mathrm{tr}( \mathbf{W}_{mk}^{H}\mathbf{W}_{mk}\bar{\mathbf{R}}_{mk} ) }\\
\end{array} \right)}
\end{align}
\setcounter{equation}{\value{mytempeqncnt2}}
\hrulefill
\vspace*{-0.45cm}
\end{figure*}


\section{Uplink Data Transmission}\label{UL}
\vspace*{-0.3cm}
In this section, we investigate the UL data transmission. Firstly, we investigate the UL achievable SE performance. Then, we design the distributed OBE combining scheme over arbitrary statistics-based channel estimators.
\vspace*{-0.3cm}
\subsection{Uplink Spectral Efficiency Analysis}\label{UL_SE}
In the UL data transmission phase, all UEs transmit their respective UL data symbols and the received UL data signal at AP $m$ can be denoted as $\mathbf{y}_m=\sum_{k=1}^K{\mathbf{g}_{mk}x_k}+\mathbf{n}_m$, where $x_k\sim \mathcal{N} _{\mathbb{C}}( 0,p_k) $ is the data symbol for UE $k$ and $\mathbf{n}_m\sim \mathcal{N} _{\mathbb{C}}( \mathbf{0},\sigma ^2\mathbf{I}_N ) $ is the UL noise at AP $m$. For UE $k$, AP $m$ selects an arbitrary receiving combining scheme $\mathbf{v}_{mk}\in \mathbb{C} ^N$ to locally decode the UL data symbol of UE $k$. Based on the fundamentals in \cite{OBETrans}, by applying the traditional use-and-then-forget (UatF) capacity bound, the achievable SE for UE $k$ under the distributed processing scheme can be computed as $\mathrm{SE}_{k}^{\mathrm{ul}}=\frac{\tau _u}{\tau _c}\log _2( 1+\mathrm{SINR}_{k}^{\mathrm{ul}} ) $ with $\mathrm{SINR}_{k}^{\mathrm{ul}}$ given as in \eqref{SINR_ul}, which is a standard result in CF mMIMO networks \cite{[162],7827017}. When applying the BE-structure combining scheme as $\mathbf{v}_{mk}=\mathbf{W}_{mk}\hat{\mathbf{g}}_{mk}$, where $\mathbf{W}_{mk}\in \mathbb{C} ^{N\times N}$ is an arbitrary channel statistics-based matrix and $\hat{\mathbf{g}}_{mk}$ is generated by arbitrary statistics-based channel estimators as in \eqref{CE}, we can derive the closed-form UL SE expressions as follows.
\vspace*{-0.3cm}

\begin{thm}\label{thm_distributed_closed}
When applying the BE-structure combining scheme $\mathbf{v}_{mk}=\mathbf{W}_{mk}\hat{\mathbf{g}}_{mk}$, the UL achievable SE expressions can be calculated in closed-form $\overline{\mathrm{SE}}_{k}^{\mathrm{ul}}=\frac{\tau _u}{\tau _c}\log _2( 1+\overline{\mathrm{SINR}}_{k}^{\mathrm{ul}} )$, where $\overline{\mathrm{SINR}}_{k}^{\mathrm{ul}}$ is given as in \eqref{SINR_Distributed_Closed}, where
\setcounter{equation}{3}
\begin{align}
&\varepsilon _{mkl}\!\!=\!\!\mathrm{tr}( \mathbf{W}_{mk}^{H}\bar{\mathbf{G}}_{mll}\mathbf{W}_{mk}\bar{\mathbf{G}}_{mkk} ) +\mathrm{tr}( \mathbf{W}_{mk}^{H}\check{\mathbf{R}}_{ml}\mathbf{W}_{mk}\bar{\mathbf{G}}_{mkk} ) \notag \\ 
&+\tau _p\mathrm{tr}( \mathbf{W}_{mk}^{H}\bar{\mathbf{G}}_{mll}\mathbf{W}_{mk}\mathbf{A}_{mk}\mathbf{\Psi }_{mk}\mathbf{A}_{mk}^{H} )\notag \\
&+\tau _p\mathrm{tr}( \mathbf{W}_{mk}^{H}\check{\mathbf{R}}_{ml}\mathbf{W}_{mk}\mathbf{A}_{mk}\mathbf{\Psi }_{mk}\mathbf{A}_{mk}^{H} ),\label{local_var}
\end{align}
$\xi _{mkl}=\sqrt{p_l}\tau _p\mathrm{tr}( \mathbf{W}_{mk}^{H}\bar{\mathbf{G}}_{mlk} ) \mathrm{tr}( \mathbf{W}_{mk}\mathbf{A}_{mk}\check{\mathbf{R}}_{ml} ) +\sqrt{p_l}\tau _p\mathrm{tr}( \mathbf{W}_{mk}^{H}\check{\mathbf{R}}_{ml}\mathbf{A}_{mk}^{H} ) \mathrm{tr}( \mathbf{W}_{mk}\bar{\mathbf{G}}_{mkl})+p_l\tau _{p}^{2}| \mathrm{tr}( \mathbf{W}_{mk}^{H}\check{\mathbf{R}}_{ml}\mathbf{A}_{mk}^{H} ) |^2$
$\chi _{mkl}=\mathrm{tr}( \mathbf{W}_{mk}^{H}\bar{\mathbf{G}}_{mlk} )$ if $l\notin \mathcal{P} _k$ and $\chi _{mkl}=\mathrm{tr}( \mathbf{W}_{mk}^{H}\mathbf{B}_{mlk} )$ if $l\in \mathcal{P} _k$, $\bar{\mathbf{G}}_{mkl}=\bar{\mathbf{g}}_{mk}\bar{\mathbf{g}}_{ml}^{H}$, $\mathbf{B}_{mlk}=\bar{\mathbf{G}}_{mlk}+\sqrt{p_l}\tau _p\check{\mathbf{R}}_{ml}\mathbf{A}_{mk}^{H}$, and 
$\bar{\mathbf{R}}_{mk}=\bar{\mathbf{G}}_{mkk}+\sqrt{p_k}\tau _p\check{\mathbf{R}}_{mk}\mathbf{A}_{mk}^{H}$.
\end{thm}
\begin{IEEEproof}
The proof can be found in Appendix~\ref{app_uplink_closed}.
\end{IEEEproof}
\vspace*{-0.4cm}
\subsection{Optimal Bilinear Equalizer Combining Design}\label{UL_OBE}
Next, we study the OBE combining design and analyze the effects of channel estimators on the achievable SE performance. Based on \eqref{SINR_ul}, we can derive OBE matrices, which can maximize the achievable UL SE as follows.
\begin{coro}\label{UL-OBE-Monte}
We can derive OBE matrices $\{ \mathbf{W}_{mk}^{*}:m=1,\dots ,M \} $, which can maximize \eqref{SINR_ul}, as $\mathbf{W}_{mk}^{*}=\mathrm{vec}^{-1}( \mathbf{w}_{mk}^{*})$, where $\mathbf{w}_{mk}^{*}\in \mathbb{C} ^{N^2}$ is the $m$-th component of $\mathbf{w}_{k}^{*}=[ \mathbf{w}_{1k}^{*,T},\dots ,\mathbf{w}_{mk}^{*,T},\dots ,\mathbf{w}_{Mk}^{*,T} ]^{T} \in \mathbb{C} ^{MN^2}$ with
\begin{equation}\label{DG_OBE_W_Monte}
\mathbf{w}_{k}^{*}\!\!=\!\!\left( \sum_{l=1}^K{p_l\mathbb{E} \{\mathbf{q}_{kl}\mathbf{q}_{kl}^{H}\}\!-\!p_k\mathbb{E} \{\mathbf{p}_k\}\mathbb{E} \{\mathbf{p}_k\}^H \!+\!\sigma ^2\mathbf{\Xi }_{k}} \right) ^{-1}\!\!\mathbb{E} \{\mathbf{p}_k\},
\end{equation}
$\mathbf{q}_{kl}=[ \mathrm{vec}( \mathbf{g}_{1l}\hat{\mathbf{g}}_{1k}^{H} ) ^T,\dots ,\mathrm{vec}( \mathbf{g}_{Ml}\hat{\mathbf{g}}_{Mk}^{H} ) ^T ]^T\in \mathbb{C} ^{MN^2}$, $\mathbf{p}_k=[ \mathrm{vec}( \mathbf{g}_{1k}\hat{\mathbf{g}}_{1k}^{H} ) ^T,\dots ,\mathrm{vec}( \mathbf{g}_{Mk}\hat{\mathbf{g}}_{Mk}^{H} ) ^T ] ^T\in \mathbb{C} ^{MN^2}$, and $\mathbf{\Xi }_{k}=\mathrm{diag}\{ ( \mathbb{E} \{ \hat{\mathbf{g}}_{1k}\hat{\mathbf{g}}_{1k}^{H} \} ^T\otimes \mathbf{I}_N ) ,\dots ,( \mathbb{E} \{ \hat{\mathbf{g}}_{Mk}\hat{\mathbf{g}}_{Mk}^{H} \} ^T\otimes \mathbf{I}_N ) \} \in \mathbb{C} ^{MN^2\times MN^2}$.
Utilizing $\{ \mathbf{W}_{mk}^{*}:m=1,\dots ,M \} $ can maximize the SINR in \eqref{SINR_ul} as 
\begin{equation}\label{SINR_max}
\begin{aligned}
&\mathrm{SINR}_{k}^{\mathrm{ul},*}=p_k\mathbb{E}\{\mathbf{p}_k\}^H\\
&\left( \sum_{l=1}^K{p_l\mathbb{E} \{\mathbf{q}_{kl}\mathbf{q}_{kl}^{H}\}\!-\!p_k\mathbb{E} \{\mathbf{p}_k\}\mathbb{E} \{\mathbf{p}_k\}^H\!+\!\sigma ^2\mathbf{\Xi }_k} \right) ^{-1}\!\!\!\!\mathbb{E} \{\mathbf{p}_k\}.
\end{aligned}
\end{equation}
\end{coro}
\begin{IEEEproof}
Note that terms of $\mathrm{SINR}_{k}^{\mathrm{ul}} $ as in \eqref{SINR_ul} can be represented as $\sum_{m=1}^M{\mathbb{E} \{\mathbf{v}_{mk}^{H}\mathbf{g}_{mk}\}}\overset{\left( a \right)}{=}\mathbf{w}_{k}^{H}\mathbb{E} \{\mathbf{p}_k\}$, $\mathbb{E} \{|\sum\nolimits_{m=1}^M{\mathbf{v}_{mk}^{H}\mathbf{g}_{ml}}|^2\}\overset{(a)}{=}\mathbb{E} \{|\sum\nolimits_{m=1}^M{\mathbf{w}_{mk}^{H}\mathrm{vec(}\mathbf{g}_{ml}\hat{\mathbf{g}}_{mk}^{H})}|^2\}=\mathbf{w}_{k}^{H}\mathbb{E} \{\mathbf{q}_{kl}\mathbf{q}_{kl}^{H}\}\mathbf{w}_k$, and
$\sum\limits_{m=1}^M{\mathbb{E} \{\parallel \mathbf{v}_{mk}\parallel ^2\}}\overset{(b)}{=}\sum\limits_{m=1}^M{\mathbf{w}_{mk}^{H}(\mathbb{E} \{\hat{\mathbf{g}}_{mk}\hat{\mathbf{g}}_{mk}^{H}\}^T\otimes \mathbf{I}_N)}\mathbf{w}_{mk}
=\mathbf{w}_{k}^{H}\mathbf{\Xi }_k\mathbf{w}_k,$
respectively, where steps (a) and (b) follow from the standard matrix computation results \cite[Eq. (34)]{OBETrans} and  \cite[Eq. (36)]{OBETrans}, respectively. Thus, we can represent \eqref{SINR_ul} as
\begin{equation}\label{UL_SINR_expansion}
\mathrm{SINR}_{k}^{\mathrm{ul}}\!\!=\!\!\frac{p_k|\mathbf{w}_{k}^{H}\mathbb{E} \{\mathbf{p}_k\}|^2}{\mathbf{w}_{k}^{H}\!(\!\sum_{l=1}^K{p_l\mathbb{E} \{\mathbf{q}_{kl}\mathbf{q}_{kl}^{H}\}\!-\!p_k\mathbb{E} \{\mathbf{p}_k\}\mathbb{E} \{\mathbf{p}_k\}^H\!\!+\!\!\sigma ^2\mathbf{\Xi }_k})\mathbf{w}_k}.
\end{equation}
Note that \eqref{UL_SINR_expansion} is a standard Rayleigh quotient to $\mathbf{w}_k$. Thus, following \cite[Corollary 2]{[162]}, we can derive the optimal $\mathbf{w}_{k}^{*}$ and corresponding maximum SINR value $\mathrm{SINR}_{k}^{\mathrm{ul},*}$.

\end{IEEEproof}
Moreover, we can obtain the closed-form expressions of $\mathbf{w}_{k}^{*}$ and $\mathrm{SINR}_{k}^{\mathrm{ul},*}$ as in Theorem~\ref{thm_distributed_obe_closed}.
\begin{thm}\label{thm_distributed_obe_closed}
We can compute $\mathbf{w}_{k}^{*}$ and $\mathrm{SINR}_{k}^{\mathrm{ul},*}$ in closed-form as $\overline{\mathbf{w}}_{k}^{*}=[ \sum_{l=1}^K{( \mathbf{\Gamma }_{kl}^{( 1 )}+\mathbf{\Gamma }_{kl}^{( 3 )} )}+\sum_{l\in \mathcal{P} _k}{( \mathbf{\Gamma }_{kl}^{( 2 )}+\mathbf{\Gamma }_{kl}^{( 4 )} )}-p_k\bar{\mathbf{r}}_k\bar{\mathbf{r}}_{k}^{H}+\sigma ^2\mathbf{\Gamma }_{k}^{( 5 )} ] ^{-1}\bar{\mathbf{r}}_k$ and $\overline{\mathrm{SINR}}_{k}^{\mathrm{ul}}=p_k\bar{\mathbf{r}}_{k}^{H}\overline{\mathbf{w}}_{k}^{*}$, respectively, where $\overline{\mathbf{r}}_k=[ \overline{\mathbf{r}}_{1k}^{T},\dots ,\overline{\mathbf{r}}_{Mk}^{T} ] ^T\in \mathbb{C} ^{MN^2}$ with $\overline{\mathbf{r}}_{mk}=\mathrm{vec}( \bar{\mathbf{R}}_{mk} ) \in \mathbb{C} ^{N^2}$, $\mathbf{\Gamma }_{kl}^{\left( 1 \right)}=\mathrm{diag}( \mathbf{\Gamma }_{1kl}^{\left( 1 \right)},\dots ,\mathbf{\Gamma }_{Mkl}^{\left( 1 \right)} ) \in \mathbb{C} ^{MN^2\times MN^2}$, $\mathbf{\Gamma }_{kl}^{\left( 2 \right)}=\mathrm{diag}( \mathbf{\Gamma }_{1kl}^{\left( 2 \right)},\dots ,\mathbf{\Gamma }_{Mkl}^{\left( 2 \right)} ) \in \mathbb{C} ^{MN^2\times MN^2}$, $\mathbf{\Gamma }_{kl}^{\left( 3 \right)}=p_l( \bar{\mathbf{G}}_{lk,\left( 2 \right)}-\bar{\bar{\mathbf{G}}}_{lk} )\in \mathbb{C} ^{MN^2\times MN^2} $, and $\mathbf{\Gamma }_{kl}^{\left( 4 \right)}=p_l( \check{\mathbf{B}}_{lk}-\bar{\mathbf{B}}_{lk}-\bar{\mathbf{G}}_{lk,\left( 2 \right)}+\bar{\bar{\mathbf{G}}}_{lk} )\in \mathbb{C} ^{MN^2\times MN^2} $, respectively. Besides, we have $\mathbf{\Gamma }_{mkl}^{\left( 1 \right)}=p_l( \bar{\mathbf{G}}_{mkk}^{T}\otimes \bar{\mathbf{G}}_{mll} ) +p_l( \bar{\mathbf{G}}_{mkk}^{T}\otimes \check{\mathbf{R}}_{ml} ) +p_l\tau _p[ ( \mathbf{A}_{mk}\mathbf{\Psi }_{mk}\mathbf{A}_{mk}^{H} ) ^T\otimes \bar{\mathbf{G}}_{mll} ] +p_l\tau _p[ \left( \mathbf{A}_{mk}\mathbf{\Psi }_{mk}\mathbf{A}_{mk}^{H} \right) ^T\otimes \check{\mathbf{R}}_{ml} ]$, $\mathbf{\Gamma }_{mkl}^{\left( 2 \right)}=p_l[ \sqrt{p_l}\tau _p\bar{\mathbf{g}}_{mlk}\tilde{\mathbf{r}}_{mlk}^{H}+\sqrt{p_l}\tau _p\tilde{\mathbf{r}}_{mlk}\bar{\mathbf{g}}_{mlk}^{H}+p_l\tau _{p}^{2}\tilde{\mathbf{r}}_{mlk}\tilde{\mathbf{r}}_{mlk}^{H} ] $
with $\tilde{\mathbf{r}}_{mlk}=\mathrm{vec}( \check{\mathbf{R}}_{ml}\mathbf{A}_{mk}^{H} )$ and $\bar{\mathbf{g}}_{mlk}=\mathrm{vec}( \bar{\mathbf{G}}_{mkl} ) $, $\bar{\mathbf{G}}_{lk,( 2 )}=\bar{\mathbf{g}}_{lk,( 2 )}\bar{\mathbf{g}}_{lk,( 2 )}^{H}$ with $\bar{\mathbf{g}}_{lk,( 2 )}=[ \bar{\mathbf{g}}_{1lk}^{T},\dots ,\bar{\mathbf{g}}_{Mlk}^{T} ] ^T$, $\bar{\bar{\mathbf{G}}}_{lk}=\mathrm{diag}( \bar{\bar{\mathbf{G}}}_{1lk},\dots ,\bar{\bar{\mathbf{G}}}_{Mlk} )$ with $\bar{\bar{\mathbf{G}}}_{mlk}=\bar{\mathbf{g}}_{mlk}\bar{\mathbf{g}}_{mlk}^{H}$, $\check{\mathbf{B}}_{lk}=\mathbf{b}_{lk}\mathbf{b}_{lk}^{H}$ with $\mathbf{b}_{lk}=\left[ \mathbf{b}_{1lk}^{T},\dots ,\mathbf{b}_{Mlk}^{T} \right] ^T$, and $\bar{\mathbf{B}}_{lk}=\mathrm{diag}( \bar{\mathbf{B}}_{1lk},\dots ,\bar{\mathbf{B}}_{Mlk} ) \in \mathbb{C} ^{MN^2\times MN^2}$ with $\bar{\mathbf{B}}_{mlk}=\mathbf{b}_{mlk}\mathbf{b}_{mlk}^{H}$ and $\mathbf{b}_{mlk}=\sqrt{p_l\tau _{p}^{2}}\mathrm{vec}( \check{\mathbf{R}}_{ml}\mathbf{A}_{mk}^{H} ) $
\end{thm}
\begin{IEEEproof}
We can easily prove it based on Theorem~\ref{thm_distributed_closed} with the aid of methods introduced in \cite[Appendix G]{OBETrans}.
\end{IEEEproof}
\begin{rem}\label{generalized}
The closed-form expressions presented in Theorem~\ref{thm_distributed_closed} and Theorem~\ref{thm_distributed_obe_closed} are generalized versions of those in \cite[Theorem 2]{OBETrans} and \cite[Theorem 4]{OBETrans}, respectively, applicable to scenarios using arbitrary statistics-based channel estimators as defined in \eqref{CE}. When $\mathbf{A}_{mk}=\sqrt{p_k}\check{\mathbf{R}}_{mk}\mathbf{\Psi }_{mk}^{-1}$, and assuming $\bar{\mathbf{G}}_{mlk} = \mathbf{0}$ due to the random phase shifts of LoS components as discussed in \cite{OBETrans}, the results in Theorem~\ref{thm_distributed_closed} and Theorem~\ref{thm_distributed_obe_closed} can reduce to those in \cite[Theorem 2]{OBETrans} and \cite[Theorem 5]{OBETrans}, respectively. The effects of $\mathbf{A}_{mk}$ can be clearly observed in  Theorem~\ref{thm_distributed_closed} and Theorem~\ref{thm_distributed_obe_closed}, which include the effects on both the LoS and NLoS component-related terms.
\end{rem}
\vspace*{-0.3cm}
\begin{rem}\label{Distributed_Implement}
We can find from Corollary~\ref{UL-OBE-Monte} and Theorem~\ref{thm_distributed_obe_closed} that the design of OBE matrices $\mathbf{W}_{mk}^{*}$ is based on global channel statistics information instead of the instantaneous one. Since the channel statistics remain constant via a long period, each AP can obtain required channel statistics via the fronthaul for each realization of AP/UE positions. Then, each AP can locally design the OBE combining schemes based on $\mathbf{W}_{mk}^{*}$ and local channel estimates, which showcase the distributed design manner. 
\end{rem}

By letting $\bar{\mathbf{g}}_{mk}=\mathbf{0}$, all results in this section are special cases to those of Rayleigh fading channels with only the NLoS component. 
When the Rayleigh fading channel is considered, the channel estimator in \eqref{CE} becomes $\hat{\mathbf{g}}_{mk}=\mathbf{A}_{mk} \mathbf{y}_{mk}^{p}$. Significantly, we derive a critical insight: when OBE combining, as outlined in \eqref{DG_OBE_W_Monte}, is applied, $\mathbf{A}_{mk}$ does not impact the maximum value of the effective SINR specified in \eqref{SINR_max}. The details are discussed in the following corollary.
\begin{coro}\label{Rayleigh_max}
When the Rayleigh fading channel is considered, the modelling of $\mathbf{A}_{mk}$ would not impact the maximum value of the effective SINR depicted in \eqref{SINR_max} with the OBE combining scheme is applied. Arbitrary $\mathbf{A}_{mk}$ would lead to a similar effective SINR value as $
\widetilde{\mathrm{SINR}}_{k}^{\mathrm{ul},*}=p_k\mathbb{E} \{\tilde{\mathbf{p}}_{k}^{H}\}(\sum_{l=1}^K{p_l\mathbb{E} \{\tilde{\mathbf{q}}_{kl}\tilde{\mathbf{q}}_{kl}^{H}\}\!-\!p_k\mathbb{E} \{\tilde{\mathbf{p}}_k\}\mathbb{E} \{\tilde{\mathbf{p}}_{k}^{H}\}\!+\!\sigma ^2\tilde{\mathbf{\Xi}}_k} ) ^{-1}$ $\mathbb{E} \{\tilde{\mathbf{p}}_k\}$, where $\tilde{\mathbf{p}}_k=[ \mathrm{vec}( \mathbf{g}_{1k}\mathbf{y}_{1k}^{p,H} ) ^T,\dots ,\mathrm{vec}( \mathbf{g}_{Mk}\mathbf{y}_{Mk}^{p,H} ) ^T ] ^T \in \mathbb{C} ^{MN^2}$, $\tilde{\mathbf{q}}_{kl}=[ \mathrm{vec}( \mathbf{g}_{1l}\mathbf{y}_{1k}^{p,H} ) ^T,\dots ,\mathrm{vec}( \mathbf{g}_{Ml}\mathbf{y}_{Mk}^{p,H} ) ^T ] ^T \in \mathbb{C} ^{MN^2}$, and $\tilde{\mathbf{\Xi}}_k=\mathrm{diag}[ ( \mathbf{y}_{1k}^{p,*}\mathbf{y}_{1k}^{p,T} ) \otimes \mathbf{I}_N,\dots ,( \mathbf{y}_{Mk}^{p,*}\mathbf{y}_{Mk}^{p,T} ) \otimes \mathbf{I}_N ] \in \mathbb{C} ^{MN^2\times MN^2}$.
\end{coro}
\begin{IEEEproof}
The proof can be found in Appendix~\ref{app_rayleigh}.
\end{IEEEproof}

Notably, for arbitrary BE matrices $\mathbf{W}_{mk}$, we can also optimize channel estimators $\mathbf{A}_{mk}$ to improve the SE performance. For the scenario with Rayleigh fading channels, $\mathbf{A}_{mk}$ can be optimized to maximize the UL SE. We define $\mathbf{a}_k=[ \mathbf{a}_{1k}^{T},\dots ,\mathbf{a}_{Mk}^{T} ]^T \in \mathbb{C} ^{MN^2}$ with $\mathbf{a}_{mk}=\mathrm{vec}( \mathbf{A}_{mk} ) \in \mathbb{C} ^{N^2}$. Following the method in  Corollary~\ref{UL-OBE-Monte}, we can derive optimal channel estimators $\mathbf{A}_{mk}^{*}=\mathrm{vec}^{-1}( \mathbf{a}_{mk}^{*} )$, where
\begin{equation}\label{Optimal_CE}
\mathbf{a}_{k}^{*}\!\!=\!\!\left( \sum_{l=1}^K{p_l\mathbb{E} \{\mathbf{z}_{kl}\mathbf{z}_{kl}^{H}\}}-p_k\mathbb{E} \{\check{\mathbf{r}}_k\}\mathbb{E} \{\check{\mathbf{r}}_k\}^H+\sigma ^2\mathbf{\Upsilon }_k \right) ^{-1}\!\!\mathbb{E} \{\check{\mathbf{r}}_k\},
\end{equation}
with the maximum UL SE value $\mathrm{SINR}_{k}^{\mathrm{ul},*}=p_k\mathbb{E} \{\check{\mathbf{r}}_k\}^H\mathbf{a}_{k}^{*}$ with $\check{\mathbf{r}}_k=[ \check{\mathbf{r}}_{1k}^{T},\dots ,\check{\mathbf{r}}_{Mk}^{T} ]^T \in \mathbb{C} ^{MN^2}$, $\check{\mathbf{r}}_{mk}=\mathrm{vec}( \mathbf{W}_{mk}^{H}\mathbf{g}_{mk}\mathbf{y}_{mk}^{p,H} )$, $\mathbf{z}_{kl}=[ \mathbf{z}_{1kl}^{T},\dots ,\mathbf{z}_{Mkl}^{T} ]^T \in \mathbb{C} ^{MN^2}$, $\mathbf{z}_{mkl}=\mathrm{vec}( \mathbf{W}_{mk}^{H}\mathbf{g}_{ml}\mathbf{y}_{mk}^{p,H} )$, and $\mathbf{\Upsilon }_k=\mathrm{diag}\{ [ \mathbb{E} \{ ( \mathbf{y}_{1k}^{p}\mathbf{y}_{1k}^{p,H} ) ^T\otimes ( \mathbf{W}_{1k}^{H}\mathbf{W}_{1k} ) \} ] ,\dots ,[ \mathbb{E} \{ ( \mathbf{y}_{Mk}^{p}\mathbf{y}_{Mk}^{p,H} ) ^T\otimes ( \mathbf{W}_{Mk}^{H}\mathbf{W}_{Mk} ) \} ] \} \in \mathbb{C} ^{MN^2\times MN^2}$, respectively. It is worth noting that when the optimal channel estimator $\mathbf{A}_{mk}^{*}$ is applied, it is not necessary to further optimize BE matrices $\mathbf{W}_{mk}$ as in Corollary~\ref{UL-OBE-Monte}. This observation follows the similar reason as in Corollary~\ref{Rayleigh_max}. Moreover, for the scenario with Rician fading channels, it is quite difficult to directly optimize $\mathbf{A}_{mk}$ due to the existence of the LoS component. Thus, the optimization of $\mathbf{A}_{mk}$ over the Rician fading channels is regarded as a vital future research direction.

\begin{table}[t!]
  \centering
  \fontsize{8}{9}\selectfont
  \caption{Computational complexity for the OBE combining scheme compared with other promising schemes with $N_r$ being the number of channel realizations.}
  \vspace*{-0.3cm}
  \label{Comparisons}
   \begin{tabular}{ !{\vrule width0.7 pt}  m{1.2 cm}<{\centering} !{\vrule width0.7pt}  m{3.1 cm}<{\centering} !{\vrule width0.7pt}  m{3.1cm}<{\centering} !{\vrule width0.7pt}}

    \Xhline{0.7pt}
         \bf Scheme & \bf Combining design  & \bf Precomputation based on statistics \cr
    \Xhline{0.7pt}
    OBE & $\mathcal{O} ( MKN^2N_r ) $   & $\mathcal{O} ( M^3KN^6+M^2KN^4\lfloor K/\tau _p \rfloor)  $ \cr\hline
    LMMSE & $\mathcal{O} ( MKN^2N_r + MN^3N_r) $  & $\mathcal{O} \left( MKN^3 \right)$\cr\hline
    LRZF & $\mathcal{O} ( MK^2NN_r + MK^3N_r) $   & $-$\cr\hline
    LSFD & $-$ & $\mathcal{O} (MK^2N^3+M^3K) $ \cr\hline
    \Xhline{0.7pt}
    \end{tabular}
  \vspace*{-0.6cm}
\end{table}

In Table~\ref{Comparisons}, we compare the computational complexity for the OBE scheme with other promising schemes, such as local MMSE (LMMSE) combining as \cite[Eq. (16)]{[162]},  local regularized zero-forcing (LRZF) combining as \cite[Eq. (4.9)]{8187178} at each AP, and the large-scale fading decoding (LSFD) scheme as \cite[Eq. (21)]{[162]} at the CPU. Computational complexity is computed for each realization of AP/UE locations. As observed, the OBE scheme involves much lower design complexity compared with the other schemes. However, the precomputation for the OBE scheme is higher than the other schemes but this complexity is acceptable since only one-time computation is required for each realization of AP/UE locations.

\newcounter{mytempeqncnt5}
\begin{figure*}[t]
\normalsize
\setcounter{mytempeqncnt5}{\value{equation}}
\setcounter{equation}{9}
\begin{equation}\label{SINR_DL_Closed}
\begin{aligned}
\overline{\mathrm{SINR}}_{k}^{\mathrm{dl}}=\frac{\left| \sum_{m=1}^M{\overline{\eta }_{mk}\mathrm{tr}\left( \mathbf{W}_{mk}^{H}\bar{\mathbf{R}}_{mk} \right)} \right|^2}{\sum_{l=1}^K{\sum_{m=1}^M{\mu _{mkl}}}+\sum_{l\in \mathcal{P} _k}{\sum_{m=1}^M{\omega _{mkl}}}+\sum_{l=1}^K{(|\sum_{m=1}^M{\lambda _{mkl}}|^2-\sum_{m=1}^M{|\lambda _{mkl}|^2})}+\sigma ^2}
\end{aligned}
\end{equation}
\setcounter{equation}{\value{mytempeqncnt5}}
\hrulefill
\vspace*{-0.6cm}
\end{figure*}

\newcounter{mytempeqncnt3}
\begin{figure*}[t]
\normalsize
\setcounter{mytempeqncnt3}{\value{equation}}
\setcounter{equation}{10}
\begin{align}
&\mathbb{E} \{ | \mathbf{v}_{mk}^{H}\mathbf{g}_{ml} |^2 \}\!=\!\underset{\left( a \right)}{\underbrace{\mathbb{E} \{ | ( \mathbf{v}_{mk}-\sqrt{p_l}\tau _p\mathbf{W}_{mk}\mathbf{A}_{mk}\mathbf{g}_{ml} ) ^H\mathbf{g}_{ml} |^2 \} }}\!+\!\underset{\left( b \right)}{\underbrace{\mathbb{E} \{ | ( \sqrt{p_l}\tau _p\mathbf{W}_{mk}\mathbf{A}_{mk}\mathbf{g}_{ml} ) ^H\mathbf{g}_{ml} |^2 \} }}\!=\!\mathrm{tr}( \mathbf{W}_{mk}^{H}\bar{\mathbf{G}}_{mll}\mathbf{W}_{mk}\bar{\mathbf{G}}_{mkk} ) \notag \\
&\!+\!\sqrt{p_l}\tau _p\mathrm{tr}( \mathbf{W}_{mk}^{H}\bar{\mathbf{G}}_{mlk} ) \mathrm{tr}( \mathbf{W}_{mk}\mathbf{A}_{mk}\check{\mathbf{R}}_{ml} ) \!+\!\tau _p\mathrm{tr}( \mathbf{W}_{mk}^{H}\bar{\mathbf{G}}_{mll}\mathbf{W}_{mk}\mathbf{A}_{mk}\mathbf{\Psi }_{mk}\mathbf{A}_{mk}^{H} ) \!+\!\sqrt{p_l}\tau _p\mathrm{tr}( \mathbf{W}_{mk}^{H}\check{\mathbf{R}}_{ml}\mathbf{A}_{mk}^{H} ) \mathrm{tr}( \mathbf{W}_{mk}\bar{\mathbf{G}}_{mkl} ) \notag \\
&+p_l\tau _{p}^{2}| \mathrm{tr}( \mathbf{W}_{mk}^{H}\check{\mathbf{R}}_{ml}\mathbf{A}_{mk}^{H} ) |^2+\mathrm{tr}( \mathbf{W}_{mk}^{H}\check{\mathbf{R}}_{ml}\mathbf{W}_{mk}\bar{\mathbf{G}}_{mkk} )+\tau _p\mathrm{tr}( \mathbf{W}_{mk}^{H}\check{\mathbf{R}}_{ml}\mathbf{W}_{mk}\mathbf{A}_{mk}\mathbf{\Psi }_{mk}\mathbf{A}_{mk}^{H} ) \label{closedmn}
\end{align}
\setcounter{equation}{\value{mytempeqncnt3}}
\hrulefill
\vspace*{-0.7cm}
\end{figure*}


\vspace{-0.4cm}
\section{Downlink Data Transmission}\label{DL}
During the phase of DL data transmission, each coherence consists of $\tau_d$ data transmissions and we consider the coherent DL transmission protocol. The DL transmitted signal from AP $m$ can be represented as $\mathbf{x}_m=\sum_{k=1}^K{\mathbf{f}_{mk}\upsilon _k}\in \mathbb{C} ^{N}$, where $\upsilon _k\sim \mathcal{N} _{\mathbb{C}}( 0,1 )$ is the DL data symbol to UE $k$ and $\mathbf{f}_{mk}\in \mathbb{C} ^{N}$ is the power scaled precoding vector for UE $k$. $\mathbf{f}_{mk}$ can be formulated as $\mathbf{f}_{mk}=\eta _{mk}\overline{\mathbf{f}}_{mk}$, where $\overline{\mathbf{f}}_{mk}$ is the precoding vector between AP $m$ and UE $k$ and $\eta _{mk}=\sqrt{{p_{mk}}/{\mathbb{E} \{ \| \overline{\mathbf{f}}_{mk} \| ^2 \}}}$ is the power factor between AP $m$ and UE $k$ with $p_{mk}$ being the power allocated for UE $k$ at AP $m$. Moreover, the transmitting power is constrained as $\sum_{k=1}^K{\mathbb{E} \{ \| \mathbf{f}_{mk} \| ^2 \}}\leqslant p_m$ with $p_m$ being the transmitting power of AP $m$. Furthermore, by assuming that the UL and DL channels are reciprocal, the received signal at UE $k$ can be denoted as 
$y_k=\sum_{m=1}^M{\mathbf{g}_{mk}^{H}\mathbf{x}_m}+n_k=\sum_{m=1}^M{\mathbf{g}_{mk}^{H}\mathbf{f}_{mk}}\upsilon _k+\sum_{l\ne k}^K{\sum_{m=1}^M{\mathbf{g}_{mk}^{H}\mathbf{f}_{ml}}\upsilon _l}+n_k$, where $n_k\sim \mathcal{N} _{\mathbb{C}}( 0,\sigma ^2) $ is the DL noise of UE $k$ with $\sigma ^2$ being the DL noise power. By applying the UatF bound \cite{8187178}, we can compute the achievable DL SE performance for UE $k$ as $\mathrm{SE}_{k}^{\mathrm{dl}}=\frac{\tau _d}{\tau _c}\log _2( 1+\mathrm{SINR}_{k}^{\mathrm{dl}}) $ with $\mathrm{SINR}_{k}^{\mathrm{dl}}$ being
\begin{equation}\label{SINR_DL}
\mathrm{SINR}_{k}^{\mathrm{dl}}=\frac{| \sum\limits_{m=1}^M{\mathbb{E} \{ \mathbf{f}_{mk}^{H}\mathbf{g}_{mk} \}} |^2}{\sum\limits_{l=1}^K{\mathbb{E} \{ | \sum\limits_{m=1}^M{\mathbf{f}_{ml}^{H}\mathbf{g}_{mk}} |^2 \} -}| \sum\limits_{m=1}^M{\mathbb{E} \{ \mathbf{f}_{mk}^{H}\mathbf{g}_{mk} \}} |^2+\sigma ^2}.
\end{equation}

By applying the channel reciprocity between the UL and DL channels, we can design $\overline{\mathbf{f}}_{mk}$ based on the UL receiving combining scheme. We define $\overline{\mathbf{f}}_{mk}=\mathbf{W}_{mk}\hat{\mathbf{g}}_{mk}$ as the DL BE precoding scheme. When the BE-structure precoding scheme is applied, the achievable DL SE expressions can be computed in closed-form as follows.
\begin{thm}\label{thm_distributed_dl_closed}
When the BE-structure precoding scheme $\overline{\mathbf{f}}_{mk}=\mathbf{W}_{mk}\hat{\mathbf{g}}_{mk}$ is applied, the achievable DL SE expressions can be computed in closed-form as $\overline{\mathrm{SE}}_{k}^{\mathrm{dl}}=\frac{\tau _d}{\tau _c}\log _2(1+\overline{\mathrm{SINR}}_{k}^{\mathrm{dl}})$, where
$\overline{\mathrm{SINR}}_{k}^{\mathrm{dl}}$ is given as in \eqref{SINR_DL_Closed} with 
$\mu _{mkl}=\bar{\eta}_{ml}^{2}\mathrm{tr}( \mathbf{W}_{ml}^{H}\bar{\mathbf{G}}_{mkk}\mathbf{W}_{ml}\bar{\mathbf{G}}_{mll} ) +\bar{\eta}_{ml}^{2}\mathrm{tr}( \mathbf{W}_{ml}^{H}\check{\mathbf{R}}_{mk}\mathbf{W}_{ml}\tilde{\mathbf{R}}_{ml} ) +\bar{\eta}_{ml}^{2}\mathrm{tr}( \mathbf{W}_{ml}^{H}\check{\mathbf{R}}_{mk}\mathbf{W}_{ml}\bar{\mathbf{G}}_{mll} ) +\bar{\eta}_{ml}^{2}\mathrm{tr}( \bar{\mathbf{W}}_{ml}^{H}\bar{\mathbf{G}}_{mkk}\mathbf{W}_{mk}\tilde{\mathbf{R}}_{ml} )$, $\omega _{mkl}=\bar{\eta}_{ml}^{2}\sqrt{p_k}\tau _p\mathrm{tr}( \bar{\mathbf{W}}_{ml}^{H}\check{\mathbf{R}}_{mk}\mathbf{A}_{ml}^{H} ) \mathrm{tr}( \bar{\mathbf{W}}_{ml}\bar{\mathbf{G}}_{mlk} ) + \bar{\eta}_{ml}^{2}\sqrt{p_k}\tau _p\mathrm{tr}( \bar{\mathbf{W}}_{ml}^{H}\bar{\mathbf{G}}_{mkl} ) \mathrm{tr}( \bar{\mathbf{W}}_{ml}\mathbf{A}_{ml}\check{\mathbf{R}}_{mk} ) +
+\bar{\eta}_{ml}^{2}p_k\tau _{p}^{2}| \mathrm{tr}( \bar{\mathbf{W}}_{ml}^{H}\check{\mathbf{R}}_{mk}\mathbf{A}_{ml}^{H} ) |^2$, $\lambda _{mkl}=\bar{\eta}_{ml}\mathrm{tr}( \mathbf{W}_{ml}\bar{\mathbf{G}}_{mlk} )$ for $l\notin \mathcal{P} _k$ and $\lambda _{mkl}=\bar{\eta}_{ml}( \mathbf{W}_{ml}\tilde{\mathbf{B}}_{mlk} )$ for $l\in \mathcal{P} _k$, $\bar{\eta}_{mk}=\sqrt{p_{mk}/[ \mathrm{tr}( \mathbf{W}_{mk}^{H}\mathbf{W}_{mk}\bar{\mathbf{G}}_{mkk} ) +\mathrm{tr}( \mathbf{W}_{mk}^{H}\mathbf{W}_{mk}\tilde{\mathbf{R}}_{mk} ) ]}$, $\tilde{\mathbf{R}}_{ml}=\tau _p\mathbf{A}_{ml}\mathbf{\Psi }_{ml}\mathbf{A}_{ml}^{H}$, and $\tilde{\mathbf{B}}_{mlk} =\bar{\mathbf{G}}_{mlk}+\sqrt{p_k}\tau _p\mathbf{A}_{ml}\check{\mathbf{R}}_{mk}$, respectively.  
\end{thm}
\begin{IEEEproof}
The proof can be easily obtained with the aid of the methods for UL as in Theorem~\ref{thm_distributed_closed} thanks to the channel reciprocity between the UL and DL channels.
\end{IEEEproof}

\begin{rem}
To enhance DL SE performance, BE matrices $\mathbf{W}_{mk}$ can be optimized during the UL data transmission phase as in Corollary~\ref{UL-OBE-Monte}. This optimization results in the OBE precoding scheme $\mathbf{f}_{mk}=\eta_{mk}\mathbf{W}_{mk}^{*}\hat{\mathbf{g}}_{mk}$. Note that the OBE precoding scheme is a heuristic one since $\mathbf{W}_{mk}$ is optimized in the UL, which can be directly applied to design the DL OBE precoding scheme. However, directly optimizing $\mathbf{W}_{mk}$ to maximize $\mathrm{SINR}_{k}^{\mathrm{dl}}$ in \eqref{SINR_DL} is very challenging due to the complex interrelationships among the $\mathbf{W}_{mk}$-related terms and the absence of a standard Rayleigh quotient relationship between $\mathrm{SINR}_{k}^{\mathrm{dl}}$ and $\mathbf{W}_{mk}$-constructed vectors. Thus, we first derive the OBE combining scheme and then apply it to the DL OBE precoding design.
\end{rem}
\vspace{-0.35cm}

\begin{rem}
Compared with \cite{OBETrans}, the technical improvements for this paper pose in the following aspects. For the uplink, the closed-form results for the achievable SE and OBE combining schemes with generalized channel estimators are derived. These important results are more generalized than the results in \cite{OBETrans}, where Remark~\ref{generalized} clarifies this insight in detail. Moreover, we investigate the achievable DL SE performance and DL OBE precoding schemes, which are not considered in \cite{OBETrans}. We also discuss many important observations, such as Corollary~\ref{Rayleigh_max}, to provide important insights for the practical implementation of the OBE beamforming schemes.
\end{rem}

\vspace{-0.7cm}
\section{Numerical Results}\label{num}
\begin{figure}[t]
\centering
\includegraphics[scale=0.45]{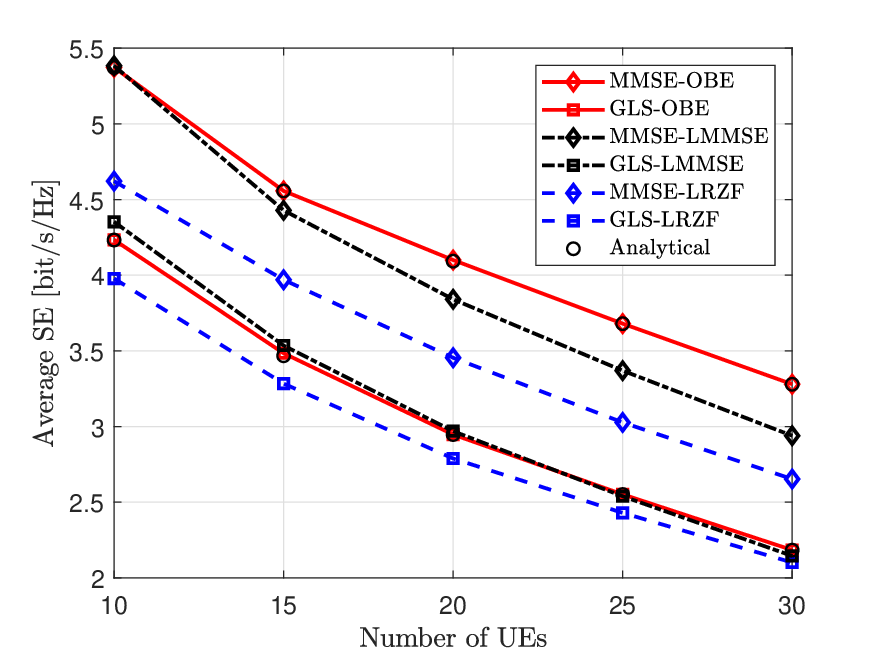}
\vspace{-0.4cm}
\caption{Average UL SE against $K$ under different combining schemes over Rician fading channels with $M=40$ and $N=4$. The legend ``A-B" denotes the ``B" combining-based scheme over the ``A" channel estimator. \label{1}}
\vspace{-0.5cm}
\end{figure}

In numerical results, all APs and UEs are randomly distributed at a $1\times 1 \, \mathrm{km}^2$ area. We assume that all APs and UEs have LoS components. We have $\tau _c=200$ and $\tau _p=1$. We consider that each coherence block is either applied for only UL or DL data transmission, which means $\tau _u=\tau _d=\tau _c-\tau _p$. In the UL, we have $p_k=200 \, \mathrm{mW}$. In the DL, for each UE, we also have $p_k=\sum_{m=1}^M{p_{mk}}=200\, \mathrm{mW}$, where the DL power is allocated based on the channel quality of UE, following the method introduced in \cite{ozdogan2019massive}. Meanwhile, we have $p_m=K\times200\, \mathrm{mW}$. Other parameters are set same as those used in \cite{OBETrans}, which are omitted due to the lack of space. In Fig.~\ref{1}, we investigate the UL SE performance against $K$ for OBE combining schemes over Rician fading channels. We consider MMSE and GLS channel estimators, and we let the two-layer processing schemes, where the LMMSE combining or LRZF combining is applied at each AP and the LSFD scheme is utilized at the CPU, as the comparison benchmarks\footnote{For the sake of fairness, we utilize the global channel statistics-based LSFD scheme at the CPU since the OBE schemes involve the global channel statistics.}. We observe that the OBE combining schemes can achieve the excellent SE performance. More specifically, the OBE schemes outperform the LMMSE-based benchmarks under the MMSE estimators and outperform the LRZF-based benchmarks under both the MMSE and GLS estimators. Moreover, over Rician fading channels, schemes based on the MMSE estimator achieve superior UL SE performance compared to those based on the GLS estimator. More importantly, the analytical results described by markers ``$\circ$" match well with the curves generated by Monte-Carlo simulations, validating our derived closed-form expressions in Theorem~\ref{thm_distributed_obe_closed}.

\begin{figure}[t]
\centering
\includegraphics[scale=0.45]{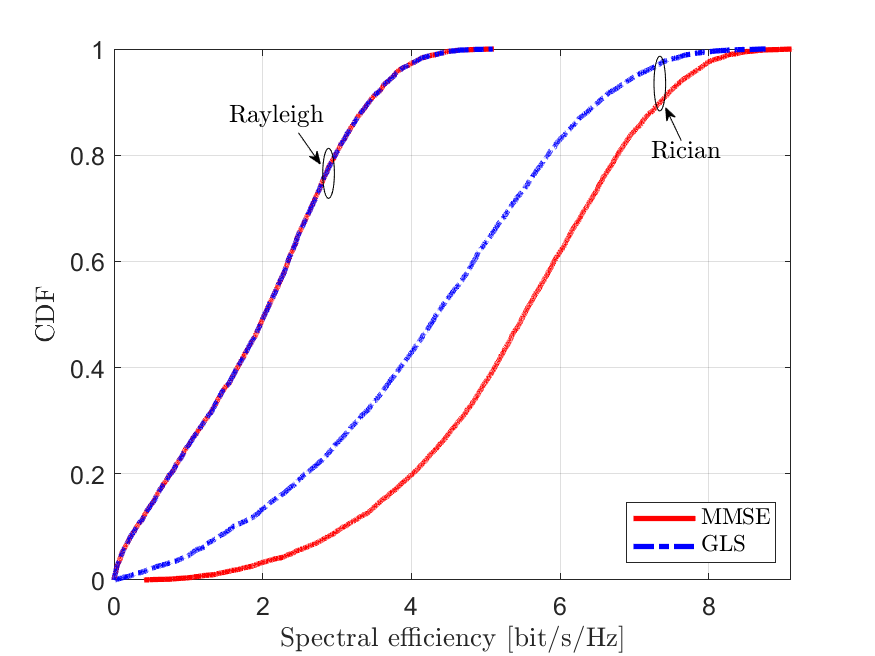}
\vspace{-0.4cm}
\caption{CDF of the UL SE per UE under OBE combining schemes over the Rician and Rayleigh fading channels with $M=40$, $K=10$, and $N=4$. \label{2}}
\vspace{-0.7cm}
\end{figure}

Fig.~\ref{2} considers the cumulative distribution function (CDF) curves of the UL SE per UE under OBE combining schemes over Rician and Rayleigh fading channels. 
We observe that, over the Rayleigh fading channel model, the GLS channel estimator can achieve the same UL SE performance as that of the MMSE channel estimator, confirming the insights discussed in Corollary~\ref{Rayleigh_max}. This observation is particularly insightful, revealing that over Rayleigh fading channels, OBE combining schemes can reduce the burden of channel estimation by applying the simplest GLS estimator since even the GLS estimator can achieve the similar SE performance to that of the computationally demanding MMSE estimator.

\begin{figure}[t]
\centering
\includegraphics[scale=0.45]{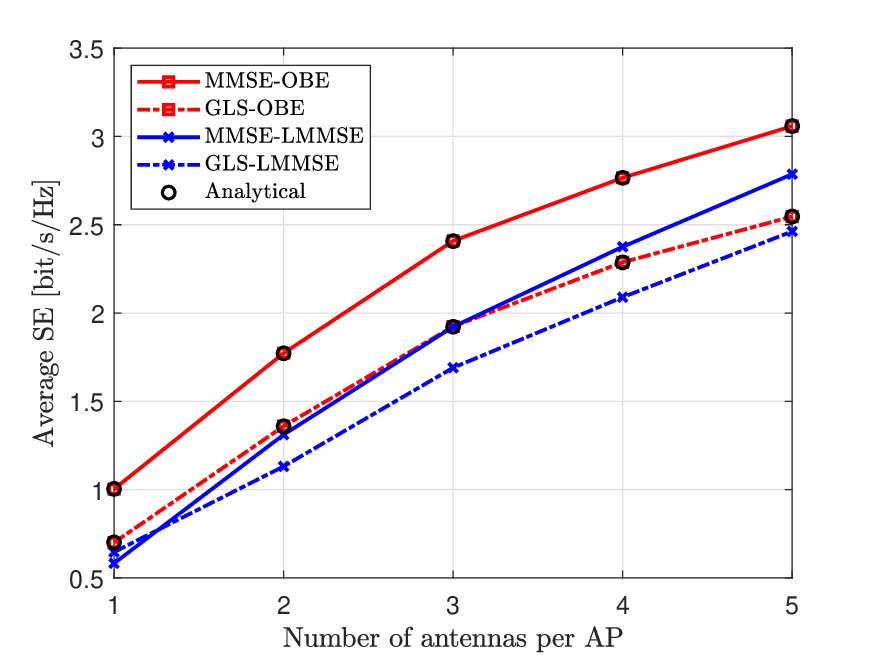}
\vspace{-0.4cm}
\caption{Average DL SE against $N$ under different precoding schemes over Rician fading channels with $M=20$ and $K=10$. The legend ``A-B" denotes the ``B" precoding-based scheme over the ``A" channel estimator.
\label{3}}
\vspace{-0.7cm}
\end{figure}

Fig.~\ref{3} shows the sum DL SE performance versus $N$ over Rician fading channels. We consider the LMMSE precoding scheme as the comparison benchmark. As observed, OBE precoding schemes can achieve better DL SE performance than that of LMMSE precoding schemes under both the MMSE and GLS channel estimators. For instance, for $N=2$, about $35 \%$ DL SE improvement can be achieved by the MMSE estimator-based OBE precoding scheme compared with the MMSE estimator-based LMMSE precoding scheme. Besides, we verify the correctness of the derived analytical results in Theorem~\ref{thm_distributed_dl_closed} by noting that ``$\circ$" generated by closed-form expressions match well with the curves generated by Monte-Carlo simulations.

\vspace{-0.5cm}
\section{Conclusions}
\vspace{-0.2cm}
We explored the OBE beamforming design for UL and DL CF mMIMO networks with arbitrary statistics-based channel estimators over Rician channels. In the UL, we derived achievable SE expressions and OBE combining schemes with arbitrary statistics-based channel estimators and their respective closed-form expressions. Notably, we observed that the achievable SE performance was not affected by the channel estimator when OBE combining schemes were applied over Rayleigh channels. In the DL, we obtained the closed-form achievable SE performance expressions with BE precoding schemes and arbitrary statistics-based channel estimators utilized. And the OBE precoding schemes were also studied with the aid of UL OBE combining. 

\vspace{-0.6cm}
\begin{appendices}

\section{Proof of Theorem~\ref{thm_distributed_closed}}\label{app_uplink_closed}
\vspace{-0.2cm}
Firstly, we have 
$\mathbb{E} \{\mathbf{v}_{mk}^{H}\mathbf{g}_{mk}\}=\mathrm{tr(}\mathbf{W}_{mk}^{H}\mathbb{E} \{\mathbf{g}_{mk}\hat{\mathbf{g}}_{mk}^{H}\})\overset{\left( a \right)}{=}\mathrm{tr(}\mathbf{W}_{mk}^{H}\bar{\mathbf{R}}_{mk})$, where step (a) is based on the independent characteristic between $\hat{\mathbf{g}}_{mk}$ and $\tilde{\mathbf{g}}_{mk}$. For $\mathbb{E} \{ | \sum_{m=1}^M{\mathbf{v}_{mk}^{H}\mathbf{g}_{ml}} |^2 \} =\sum_{m=1}^M{\sum_{n=1}^M{\mathbb{E} \{ ( \mathbf{v}_{mk}^{H}\mathbf{g}_{ml} ) ^H( \mathbf{v}_{nk}^{H}\mathbf{g}_{nl} ) \}}}$, there are four possible cases for the AP-UE combinations. Taking the scenario with $m=n,\ l\in \mathcal{P} _k$ as an example, $\mathbf{v}_{mk}$ and $\mathbf{g}_{ml}$ are dependent, we can rewrite $\mathbf{v}_{mk}$ as $\mathbf{v}_{mk}=\mathbf{v}_{mk}-\sqrt{p_l}\tau _p\mathbf{W}_{mk}\mathbf{A}_{mk}\mathbf{g}_{ml}+\sqrt{p_l}\tau _p\mathbf{W}_{mk}\mathbf{A}_{mk}\mathbf{g}_{ml}$, where $\mathbf{v}_{mk}-\sqrt{p_l}\tau _p\mathbf{W}_{mk}\mathbf{A}_{mk}\mathbf{g}_{ml}$ and $\mathbf{g}_{ml}$ are independent. Thus, we can compute $\mathbb{E} \{ ( \mathbf{v}_{mk}^{H}\mathbf{g}_{ml} ) ^H( \mathbf{v}_{nk}^{H}\mathbf{g}_{nl} ) \} =\mathbb{E} \{ | \mathbf{v}_{mk}^{H}\mathbf{g}_{ml} |^2 \} $ as \eqref{closedmn}, where term (a) and term (b) in \eqref{closedmn} are computed based on \cite[Lemma 4]{ozdogan2019massive} and \cite[Lemma 5]{ozdogan2019massive}, respectively. The computation of other AP-UE combinations can be easily derived based on the similar method and is therefore omitted due to the lack of space. Finally, for $\mathbb{E} \{ \| \mathbf{v}_{mk} \| ^2 \}$, we have $\mathbb{E} \{\parallel \mathbf{v}_{mk}\parallel ^2\}=\mathrm{tr}( \mathbf{W}_{mk}\mathbb{E} \{\hat{\mathbf{g}}_{mk}\hat{\mathbf{g}}_{mk}^{H}\}\mathbf{W}_{mk}^{H} ) =\mathrm{tr(}\mathbf{W}_{mk}^{H}\mathbf{W}_{mk}\bar{\mathbf{R}}_{mk})$. Combining all derived results, we can derive the closed-form UL achievable SE expressions in Theorem~\ref{thm_distributed_closed}.

\vspace{-0.8cm}
\section{Proof of Corollary~\ref{Rayleigh_max}}\label{app_rayleigh}
\vspace{-0.2cm}
We have $\hat{\mathbf{g}}_{mk}=\mathbf{A}_{mk} \mathbf{y}_{mk}^{p}$ for Rayleigh channels. By substituting $\hat{\mathbf{g}}_{mk}$ into \eqref{SINR_max}, we have $\mathrm{vec}( \mathbb{E} \{ \mathbf{g}_{mk}\hat{\mathbf{g}}_{mk}^{H} \} ) =\mathrm{vec}( \mathbb{E} \{ \mathbf{g}_{mk}\mathbf{y}_{mk}^{p,H} \} \mathbf{A}_{mk}^{H} ) \overset{\left( a \right)}{=}( \mathbf{A}_{mk}^{*}\otimes \mathbf{I}_N ) \mathrm{vec}( \mathbb{E} \{ \mathbf{g}_{mk}\mathbf{y}_{mk}^{p,H} \} ) $, where step (a) follows from $\mathrm{vec}( \mathbf{XYZ} ) =( \mathbf{Z}^T\otimes \mathbf{X} ) \mathrm{vec}( \mathbf{Y} ) $.
We can further derive
$\mathbb{E} \{\mathbf{p}_k\}=\{ [ ( \mathbf{A}_{1k}^{*}\otimes \mathbf{I}_N ) \mathrm{vec(}\mathbb{E} \{\mathbf{g}_{1k}\mathbf{y}_{1k}^{p,H}\}) ] ^T,\dots,
[ ( \mathbf{A}_{Mk}^{*}\otimes \mathbf{I}_N ) \mathrm{vec(}\mathbb{E} \{\mathbf{g}_{1k}\mathbf{y}_{1k}^{p,H}\}) ] ^T \} ^T=\mathrm{diag}[ ( \mathbf{A}_{1k}^{*}\otimes \mathbf{I}_N ) ,\dots ,( \mathbf{A}_{Mk}^{*}\otimes \mathbf{I}_N ) ]\mathbb{E}\{\tilde{\mathbf{p}}_k\}$. Similarly, we can derive
$\mathbb{E} \{\mathbf{p}_k\}^H=\mathbb{E}\{\tilde{\mathbf{p}}_k\}^H\mathrm{diag}[ ( \mathbf{A}_{1k}^{T}\otimes \mathbf{I}_N ) ,\dots ,( \mathbf{A}_{Mk}^{T}\otimes \mathbf{I}_N ) ]$,
by applying  $( \mathbf{X}\otimes \mathbf{Y} ) ^H=\mathbf{X}^H\otimes \mathbf{Y}^H$. We define $\mathbf{A}_{k}=\mathrm{diag}[ ( \mathbf{A}_{1k}^{T}\otimes \mathbf{I}_N ) ,\dots ,( \mathbf{A}_{Mk}^{T}\otimes \mathbf{I}_N ) ]$.
Furthermore, we can easily derive $\mathbb{E}\{\mathbf{q}_{kl}\mathbf{q}_{kl}^{H}\}=\mathbf{A}_{k}^{H}\mathbb{E}\{\tilde{\mathbf{q}}_{kl}\tilde{\mathbf{q}}_{kl}^H\}\mathbf{A}_{k}$. $\tilde{\mathbf{p}}_{k}$ and $\tilde{\mathbf{q}}_{kl}$ are defined in Corollary ~\ref{Rayleigh_max}. As for $\mathbf{\Xi }_{k}$, we have
$\mathbb{E} \{\hat{\mathbf{g}}_{mk}\hat{\mathbf{g}}_{mk}^{H}\}^T\otimes \mathbf{I}_N=( \mathbf{A}_{mk}^{*}\mathbb{E} \{\mathbf{y}_{mk}^{p,*}\mathbf{y}_{mk}^{p,T}\}\mathbf{A}_{mk}^{T} ) \otimes \mathbf{I}_N\overset{( a )}{=}( \mathbf{A}_{mk}^{*}\otimes \mathbf{I}_N ) \mathbb{E} \{[ ( \mathbf{y}_{mk}^{p,*}\mathbf{y}_{mk}^{p,T} ) \otimes \mathbf{I}_N ] \}( \mathbf{A}_{mk}^{T}\otimes \mathbf{I}_N )$,
where step (a) follows from $ \mathbf{UY}\otimes \mathbf{XZ}=( \mathbf{U}\otimes \mathbf{X} ) ( \mathbf{Y}\otimes \mathbf{Z} )$. Thus, we can represent $\mathbf{\Xi }_{k}$ as $\mathbf{\Xi }_{k}=\mathbf{A}_{k}^{H}\mathrm{diag}\mathrm{[(}\mathbf{y}_{1k}^{p,*}\mathbf{y}_{1k}^{p,T})\otimes \mathbf{I}_N,\dots ,(\mathbf{y}_{Mk}^{p,*}\mathbf{y}_{Mk}^{p,T})\otimes \mathbf{I}_N]\mathbf{A}_k$. In summary, by substituting these terms into \eqref{SINR_max} and applying $( \mathbf{XYZ} ) ^{-1}=\mathbf{Z}^{-1}\mathbf{Y}^{-1}\mathbf{X}^{-1}$, we can reformulate $\mathrm{SINR}_{k}^{\mathrm{ul},*}$ in \eqref{SINR_max} as $\widetilde{\mathrm{SINR}}_{k}^{\mathrm{ul},*}$.

\end{appendices}

\bibliographystyle{IEEEtran}
\vspace{-0.6cm}
\bibliography{IEEEabrv,Ref}

\end{document}